
\nonstopmode
 \tolerance=10000
\input phyzzx.tex


\def\np{Nucl. Phys.}
\def\pl{Phys. Lett.}

\def\intmp{Intern. J. Mod. Phys.}


\REF\thierry{ J. Thierry-Mieg,
{\it `BRS analysis of Zamolodchikov spin 2 and 3 current algebra'},
\pl\ {\bf B197} (1987) 368.}

\REF\noncri{M. Bershadsky, W. Lerche, D. Nemeschansky and N. P. Warner,
{\it `A BRST operator for the non-critical \hbw-strings'},
\pl\ {\bf B292} (1992) 35.}

\REF\bergss{E. Bergshoeff, A. Sevrin and X. Shen,
{\it `A derivation of the BRST operator for the non-critical \hbw-strings'},
\pl\ {\bf B296} (1992) 95.}

\REF\lupwa{H. Lu, C. N. Pope and X. J. Wang,
{\it `On higher-spin generalisation of string theory'},
\intmp\ {\bf A9} (1994) 1527.  hep-th/9304115 .}

\REF\lupwaa{H. Lu, C. N. Pope and X. J. Wang and S. C. Zhao,
{\it `Critical and non-critical $\hbw_{2,4}$ strings'},
Class. Q. Grav {\bf 11} (1994) 939. hep-th/9311084.}

\REF\lupwaaa{H. Lu, C. N. Pope and X. J. Wang and S. C. Zhao,
{\it `A note on $\hbw_{2,s}$ strings'},
\pl\ {\bf B237} (1994) 261.
hep-th/9402133.}

\REF\bowk{ P. Bouwknegt,
{\it `Extended conformal algebras'},
\pl\ {\bf B207} (1988) 295.}

\REF\hama{K. Hamada and M. Takao,
{\it `Spin-4 current algebra'},
\pl\ {\bf B209} (1988) 247.}

\REF\zhang{ D. Zhang,
{\it `Spin-4 extended conformal algebra'},
\pl\ {\bf B232} (1989) 323.}

\REF\nah{R. Blumenhagen, M. Flohr, A. Kliem, W. Nahm, A. Recknagel and
 R. Varnhagen,
{\it `\hbw-algebras with two and three generators'},
\np\ {\bf B361} (1991) 255.}

\REF\kawat{H. G. Kausch and G. M. T. Watts,
{\it `A study of \hbw-algebras using Jacobi identities'},
\np\ {\bf B354} (1991) 740.}

\REF\vafa{ C. Vafa,
{\it `Toward classification of conformal theories'},
\pl\ {\bf B206} (1988) 421.}

\REF\berg{E. Bergshoeff, H.J. Boonstra, M. de Roo, S. Panda
and A. Sevrin,
{\it `On the BRST operator of \hbw-strings'},
\pl\ {\bf B308} (1993) 34-41.}

\REF\qmwbrst{J. M. Figueroa-O'Farrill, C. M. Hull, L. Palacios and
E. Ramos, {\it `Generalised $\hbw_3$-strings from free fields'},
QMW preprint QMW-PH-93-34 (1994), hep-th/9409129.}

\REF\thiel{K. Thielemans,
{\it `A mathematica package for computing OPE'},
\intmp\ {\bf C2} (1992) 787.}


\def\W{{\cal W}}
\def\half{{\textstyle {1 \over 2}}}
\def\hbw {\hbox{\W}}
\def\wtf{$\W_{2,5}$}
\def\wtc{$\W_{2,4}$}
\def\wts{$\W_{2,s}$}
\def\cto{c_t}

\def\tz{T(z)}
\def\tw{T(w)}

\def\tpw{T'(w)}

\def\tiz#1#2{T_{#1}(#2)}
\def\tipz#1#2{T_{#1}'(#2)}

\def\dtiz#1#2#3{T_{#1}^{(#2)}(#3)}

\def\zw{z - w}

\def\wpdew{W'(w)}

\def\wdew{W(w)}
\def\wdez{W(z)}
\def\dwz#1#2{W^{(#1)}(#2)}

\def\wipz#1#2{W_{#1}'(#2)}


\def\ooz {\over {(z-w)}^}

\def\cww{C_{ww}^w}


\Pubnum = {IC/94/336}
\date = {Sept 1994}
\pubtype={hep-th/9410224}

\titlepage
\title {\bf A BRST charge for non-critical $\W_{2,s}$
strings}

\author {L. Palacios}


\
\address {International Centre for Theoretical Physics
\break
P.O. Box 586, 34100 Trieste,\break
Italy }


\abstract
{We present a general argument for the construction of
BRST charges of the `non-critical' $\W_{2,4}$, $\W_{2,5}$, $\W_{2,6}$,
and $\W_{2,8}$ strings.
This evidences the existence of BRST charges
for a kind of soft-type algebras
which can be constructed from two copies of quantum $\W_{2,s}$
algebras, (s=3,4,5,6,8).}

\endpage
\pagenumber=1
\chapter {Introduction}
Much work has been done on the construction and analysis of
the BRST charges
for \W\ algebras since the construction of the nilpotent
BRST charge for the $\W_3$ algebra by Thierry-Mieg [\thierry].
An interesting BRST operator
is the one associated with a direct product of two quantum
$\W_3$ algebras [\noncri].
This direct product does not form a $\W_3$ algebra but a soft-type
algebra.
This result can also be interpreted as the BRST charge for
the $\W_3$ matter coupled to $\W_3$-gravity, extending in this way the
theory of two-dimensional gravity.
In this construction, one copy of the $\W_3$ algebra corresponds
to the matter system and the other to the $\W_3 $ gravity sector
that can be represented by an SL3 Toda field theory.
It has also been proved that this BRST charge  appears in the quantisation
of some classical soft-type algebras [\bergss].
Due to the non-linear properties of the $\W$ algebra, this BRST charge
provides a basis for study of a nontrivial extension of the
cohomology of the $\W_3$ algebra.

Recently, some nilpotent BRST operators have been found with no
reference as
to whether the quantum algebra exists or not [\lupwa, \lupwaa].
In this construction, it is possible to obtain several nilpotent
BRST operators
using spin-2 generators and a spin-s generator formed from
a free field and ghost fields. These BRST charges
are usually referred to as the BRST operators of
the $\W_{2, s}$ strings.
This construction gives the BRST charge as the sum of two commuting
BRST operators.
Using certain indirect methods,
it has been possible
to relate these
BRST operators to some quantum algebras, for instance, the  two
BRST charges for the case $s=4$ are  associated with
quantum $\W B_2$ algebra, two of the four $s=6 $ BRST operators
can be associated with quantum $\W G_2$ algebra [\lupwaaa].
The cohomology related to these nilpotent operators has been
discussed
in  [\lupwaa, \lupwaaa].

In the case $s=5$, it seems that there would appear to be only one
nilpotent  BRST operator and the nilpotency condition
requires a value of the central charge of $\cto = 268 $. On the other
hand, a quantum \wtf\ algebra that
could be related to this BRST charge is consistent for a discrete set
of values
of the central charge that does not include 268. A question
that can be raised is
what kind of algebraic structure is behind the quantum BRST
operator constructed in [\lupwa].

In the case of the \wtf\ algebra, we should  notice that the value of the
central charge $(\cto = 268) $, required by the nilpotency of the
BRST operator, is not
included in the discrete set of values for which a quantum $\W_{2, 5}$
algebra  is consistent. However, an interesting fact is that
there are two
particular values of the central charge,
$c_{\pm} = 268/2 \pm 60 \sqrt{5}$, for
which the quantum $\W_{2,5}$ algebra is consistent. Furthermore,
for these values of the central charge, the Jacobi identities
are satisfied in a consistent way.
Then, in an attempt to identify an underlying algebraic
structure  of the $s=5$ BRST operator and using this apparent
coincidence,
one is tempted to consider the possibility of constructing a
BRST charge associated with two copies of the quantum \wtf\ algebras
in similar fashion as has been done for the $\W_3$ algebra [\noncri].

Here, we present the
BRST operator of some classical non-linear algebras and the explicit
expression for a BRST charge associated
with \wtc\ $\times$ \wtc. The left  and right
sectors are consistent quantum algebras at
$ c_{\pm} = \half c_t \pm 60 \sqrt{2} $, where $\cto = 4( 3s^2 - 3 s + 7 )$
is the contribution of the ghost system $\{ b, c \}$ and
$\{ \beta, \gamma \}$ with conformal weights $(2, -1)$ and
$(s, 1-s)$ respectively.
Similar irrational values of the central charge occur in
the quantum \wts\ algebras (s= 5, 6, 8).
Then, a similar construction might be possible for these
cases. This will be presented elsewhere.

\chapter {BRST charges for some  classical algebras.}
First, we shall consider the BRST operator associated with the following
classical algebras written in an OPE formalism
$$\eqalign{ \tz \tw &= {2\tw \ooz2} + {\tpw \over\zw}\cr
            \tz \wdew &= {s \wdew\ooz2} + {\wpdew \over\zw}\cr
            \wdez \wdew &= {2 \kappa^2 T^{s-1}(w)\ooz2} +
            { \kappa^2 \partial T^{s-1}(w) \over\zw}\cr}\eqn\clasw$$
The generators $T$ and $W$ have spin two and three respectively.
Starting with a general Ans\"atze for the classical BRST charge and requiring
the nilpotency of  this operator one obtains
$$ Q = Q_0 + Q_1 + Q_2\eqn\claqu$$
where
$$ Q_0 (s) = \oint c(T + c' b + s \gamma ' \beta + (s-1) \gamma '
\beta)\eqn\qzero$$
$$ Q_1 (s) = \oint \gamma W \eqn\qune$$
$$ Q_2 (s) = \oint \kappa^2 \gamma \gamma ' T^{s-2} b\eqn\qdue$$

We have introduced the ghost-antighost systems $(c,b)$ and $(\gamma, \beta)$
of spin $(-1, 2)$ and $(1-s, s)$
associated with the generators $T$ and $W$ respectively.
We should notice that \qdue\ contains the
field-dependent  structure coefficients of the algebra.

Next, we consider two commuting copies of the algebra \clasw\ and define
the following generators
$$ T = T_1 + T_2 \eqn\tplust$$
$$ W = {W_1\over \kappa_1} + i^{s -4} {W_2\over \kappa_2}
\eqn\wplusw$$
Assuming the most general form for a BRST charge of the constraints
\tplust\ and \wplusw, the existence of
the algebras of the type \clasw\ and requiring  nilpotency for a BRST
operator $Q$, we obtain
$$ Q = Q_0 + Q_1 + Q_2\eqn\clasww$$
where
$$ Q_0 (s, s)= \oint c(T + c' b + s \gamma ' \beta + (s-1) \gamma
\beta)\eqn\qwwzero$$ $$ Q_1(s,s) = \oint \gamma W \eqn\qwwune$$
$$ Q_2(s, s) = \oint \gamma \gamma' (T_1^{s-2} - T_1^{s-3} T_2 +
\ldots - T_2^{s-2})b\eqn\qwwdue$$
On the other hand, the soft-type algebra that arises from
the two copies of
\clasw\ is the following
$$\eqalign{ \tz \tw &= {2\tw \ooz2} + {\tpw \over\zw}\cr
            \tz \wdew &= {s \wdew\ooz2} + {\wpdew \over\zw}\cr
            \wdez \wdew &= {2 F(T_1, T_2) T\ooz2} +
            {\partial F(T_1, T_2) T \over\zw}\cr}\eqn\clasww$$
where $F(T_1, T_2) = T_1^{s-2} - T_1^{s-3} T_2 + \ldots - T_2^{s-2}$.
The BRST charge for this non-linear algebra is also given by \claqu, and
it can be obtained following the standard method for the
construction of a classical BRST charge.

We thus see that at the classical level it was possible to associate a
BRST charge with two copies of classical non-linear algebras
after only requiring nilpotency to a general Ansatz of a BRST operator that
contains generators of a non-linear classical algebra.

We would like to extend this straightforward method for the construction
of a nilpotent BRST operator that will contain generators of
consistent quantum algebras at certain values of the central charge,
as to
compute $Q^2$ we will need the OPE's  of the generators
within the general Ansatz.

In the next sections, we shall present a concrete example of this idea, the
result can be seen as the quantisation of the kind of classical BRST
operators presented in this section.
\vfill\eject

\chapter{The quantum \wts\ algebras.}
It is well known by now that the quantum \wts\ $( s= 3, 4, 5, 6,7, 8) $
algebras have  the general
form [\bowk, \nah, \kawat]
$$\eqalign{ \tz \tw &= {c/2\ooz4} +  {2\tw \ooz2} + {\tpw \over\zw}\cr
            \tz \wdew &= {s \wdew\ooz2} + {\wpdew \over\zw}\cr
            \wdez \wdew &= (c/s) [{\bf 1}] + C_{ww}^w [W] \cr}\eqn\wavecw$$
where $\tz$ is the spin-2 field generating a Virasoro algebra with central
charge $c$ and $\wdez$ is a spin-s primary field.
When the spin-s is an odd number, the coefficient $C_{ww}^w$
must be equal to zero.
Several results have shown that the algebras in some cases are consistent for
generic values of the central charge but in other cases the algebras
are consistent only for a discrete set of values of the central
charge
[\bowk, \nah, \kawat].
For example, \wtc\ is consistent for  $c = 86 \pm 60 \sqrt{2}$
with $C_{ww}^w= 0$, and for generic values of $c$ with $ \cww \neq 0$
(except some particular values of $c$
that can be read off from the coefficients of the algebra)
[\bowk, \hama, \zhang, \nah, \kawat].
This algebra is related to the $\W B_2$ algebra.
For $s=5$, the \wtf\ algebra is consistent only for a discrete set of
values of
the central charge, namely $ c= 6/7, -350/11, -7$ and some curious values
$ c = 134 \pm 60 \sqrt{5}$. The first set are rational c-values which
can be obtained from the minimal series of $E(6)$ Casimir algebra, from
which one can obtain the \wtf\ as a contraction algebra at these
rational values of the central charge.
The second set contains irrational values of the central charge.
These values
would apparently not correspond to RCFT's based on the
result that states that
having
a finite set of primary fields in the operator content of a CFT implies the
rationality of the central charge [\vafa].
{}From \wavecw, we could use the quantum algebras with $s=4, s=5$
and $s=6$
at the irrational values of the central charge and
construct a BRST charge associated  with two copies of these
quantum algebras which are the
consistent quantum versions  of the algebras
\clasw.

The details of OPE \wavecw\ for the  quantum \wtc\  algebra are given below
$$\eqalign{\wdez \wdew
          &= {c/4\ooz8} + {2\tiz{}{w}\ooz6} + {\tipz{}{w} \ooz5}\cr
          &+ {1\ooz4}( b_{1} \Lambda + b_{2} \wdew + 3/10
\dtiz{}{2}{w})\cr
  & + {1\ooz3} (1/2 b_{1}\Lambda' + 1/2 b_{2}\wipz{}{w} + 1/15
\dtiz{}{3}{w})\cr
  &+ {1\ooz2} (5/36 b_{1}\Lambda ^{(2)}
   + 5/36 b_{2}\dwz{2}{w} + b_{3}P + b_{4}D \cr
 &\qquad \qquad + b_{5} A(w) + 1/84 \dtiz{}{4}{w})  \cr
 &+ {1\over\zw}(1/36 b_{1}\Lambda ^{(3)}
+ 1/36 b_{2}\dwz{3}{w} + 1/2 b_{3} P' + 1/2 b_{4}D' \cr
&\qquad \qquad + 1/2 b_{5} A'(w) + 1/560 \dtiz{}{5}{w}) \cr} \eqn\qwcc$$
where the coefficients and the composite fields are
$$\Lambda = (TT) - 3/10 T''\eqn\avier$$
$$P = (TT'') - (TT')' + 2/9 (TT)'' - 1/42 T^{(4)}\eqn\asich$$
$$D = (T\Lambda) - 1/6 \Lambda '' \eqn\asichh$$
$$A = (T W) - 1/6 W'' \eqn\aeven$$
$$\eqalign{ b_{1} &= {42\over 22 + 5 c}\cr
   b_{2}&= \sqrt{{54(24 + c)(196 - 172 c + c^2)}\over
     (-1 + 2 c)(22 + 5 c) (68 + 7 c)}\cr
   b_{3} &= {3 (-524 + 19 c)\over 10 (-1 + 2 c)(68 + 7 c)}\cr
b_{4} &= {24 (13 + 72 c)\over (-1 + 2 c)(22 + 5 c)(68 + 7 c)}\cr
b_{5} &= {28 b_{2} \over 3(24 + c)}\cr}\eqn\wccoe$$
in this case we are
considering the values of the central charge
$c_{\pm } = 86 \pm 60 \sqrt{2}$
for which $b_{2} = 0 =  b_{5} $.

\chapter{The quantum BRST charge}
First, we can give some generalities of the BRST charge
associated with a quantum \wts\ algebra \wavecw.
The most general form of the BRST charge can in principle be written as
the sum of $s$ BRST charges, $Q_n(s)$, each one with a ghost number
and conformal weight equal to one and a $(\beta - \gamma)$-ghost-number
equal to $n$ [\qmwbrst]
$$ Q(s) = Q_0(s) + Q_1(s) + \ldots + Q_{s-1}(s)$$
At the quantum level, the classical result
$Q_n(s) = 0 $ for $n \geq 3$ would appear to be preserved
[\berg, \lupwa, \qmwbrst]. The nilpotency of this BRST charge is
then equivalent to the following
$$Q_0^2 = 0\eqn\nilqzero$$
          $$ [Q_0, Q_1] = 0 \eqn\nilqone$$
         $$  [Q_0, Q_2] + \half [Q_1, Q_1] = 0\eqn\nilqtwo$$
The characteristics of $Q_0$ and the condition \nilqone\ result in the
following form for $Q_0$
$$Q_0 =\oint c(T + \half T_{bc} + T_{\beta \gamma})\eqn\qqzero$$
$$T_{bc} = - 2 bc' - b' c \eqn\tbc$$
$$T_{\beta \gamma}= - s \beta \gamma' - (s - 1) \beta ' \gamma  \eqn\bega$$
$Q_0$ in \qqzero\ satisfies the nilpotency condition  provided that
the central charge is given by $$ c = 4( 3 s^2 - 3 s + 7)\eqn\ccharge$$
The form of $Q_1$ is given by
$$Q_1 = \oint \gamma W(z)\eqn\qqune $$
Since \wdez\ is a primary field, the condition
$[Q_0, Q_1] = 0$ will automatically be satisfied and any possible extra
term has to be equal to zero [\qmwbrst].
These results can also be applied to BRST operators associated with
\wts\ $\times$ \wts\ algebras, but with the generators $ T$ and $W$
defined similarly as in \tplust\ and \wplusw. Then
$$Q_1(4, 4) = \oint \gamma [{W_1(z)\over \sqrt{b_{4+}/2}}
+ {W_2(z)\over \sqrt{b_{4-}/2}}]\eqn\cuaqqune$$
where $b_{4\pm}$ are fixed by the condition \nilqtwo\ and coincide
with \wccoe\ evaluated at $c_{\pm } = 86 \pm 60 \sqrt{2}$.

 In a previous section, we have presented
the general form for the
BRST charge associated with classical algebras of the type
\clasw\ and \clasww. In particular, the term
$Q_2(s, s)$ associated with two classical algebras \clasw\ is given in
\qwwdue.
Here we present the quantum version for the case of $(4, 4)$,
that in principle is of the form
$$ Q_2(s,s) = \oint \gamma \gamma' (T_1^{s-2} - T_1^{s-3} T_2 +
\ldots T_2^{s-2})b + O(\hbar^k )\eqn\qqwwdue$$
In the case of the BRST operator for \wtc\ $\times$  \wtc,
we have obtained that indeed $Q_3 = 0$ and $Q_2$ is given as follows\footnote
\dagger{We have used the mathematica program in [\thiel]}
$$ \eqalign{Q_2(4, 4) &= \oint \gamma \gamma'
                [(T_1 T_1) - T_1 T_2 +  (T_2 T_2)] b\cr
               &+\gamma\gamma'b[{-7\over 4}(T_1 + T_2)''
                + 2\sqrt{2}(T_1 - T_2)''] \cr
               &+ \gamma\gamma^{(3)}b [{7\over 4}(T_1 + T_2)
               - {7\sqrt{2}\over 2} (T_1 - T_2)]\cr
               &+ \gamma\gamma^{(2)}b' [{5 \over2}(T_1 + T_2)
               - {11\sqrt{2}\over 2} (T_1 - T_2)] \cr
               &+ \gamma\gamma'b^{(2)} [{7\over 4}(T_1 + T_2)
               -{7\sqrt{2}\over 2}(T_1 - T_2)] \cr
               &- {7\over 48}(45 \gamma\gamma^{(5)}b -
                   76 \gamma\gamma^{(3)}b^{(2)}
                 - 45 \gamma\gamma^{(2)}b^{(3)}) \cr
              &- {7\over 4}( \gamma\gamma^{(2)}c'b b'
               + \gamma\gamma 'c'b b^{(2)}) \cr
               &+ {7\over 4} \gamma\gamma '\gamma^{(2)}b \beta '
             \cr}\eqn\cuaqqdue$$
We have started with a general Ansatz for $Q_2$ and imposed the
condition \nilqtwo\ to  fix the coefficients of the Ansatz. We should also
notice that only the OPE WW \qwcc\
with arbitrary coefficients $b_i$ is needed and these will be fixed
by \nilqtwo\ and given by \wccoe\ evaluated at $c_{\pm } = 86 \pm 60
\sqrt{2}$.
Finally, a BRST charge associated with \wtc\
$\times$ \wtc\ is given by \qqzero, \cuaqqune\ and \cuaqqdue.

\chapter{Conclusions}
These results have shown the existence of
a BRST charge associated with an algebraic
structure \wtc\ $\times$ \wtc, which is of a different nature as the
ones constructed in [\lupwaa]. It would be interesting to see their
connection.
Our construction has been possible due to the existence
of quantum non-linear algebras at
particular irrational values of
the central charges of the form $ c = \half \cto  \pm \sqrt{k}$, where
$\cto $ is the value of the central charge needed to construct a
nilpotent BRST charge.
These values occur for the \wts\ $(s= 4, 5, 6, 8)$ and it would
be interesting
to investigate further the meaning of these special values and in
particular, the existence of similar BRST charges.

Acknowledgments

The author would like to thank C. N. Pope for useful discussions and
S. Randjbar-Daemi for his support.
I would also like to thank Professor Abdus Salam, the International
Atomic Energy Agency and UNESCO for hospitality at the International
Centre for Theoretical Physics, Trieste.

\refout
\bye
\end